\documentclass[12pt]{article}

\usepackage[margin=1in]{geometry}
\usepackage{microtype}
\usepackage{hyperref}
\usepackage{url}
\usepackage{booktabs}
\usepackage{graphicx}
\usepackage{multirow}
\usepackage{amsmath}
\usepackage{subcaption}
\usepackage{natbib}          

\title{Polarization by Default: Auditing Recommendation Bias in LLM-Based Content Curation}

\author{
  Nicolò Pagan\thanks{University of Zurich, Department of Informatics, Zurich, CH \texttt{nicolo.pagan@uzh.ch}} \and
  Christopher Barrie\thanks{New York University, Department of Sociology, New York, USA and University of Oxford, Department of Sociology, Oxford, UK} \and
  Chris A. Bail\thanks{Duke University, Department of Sociology, Computer Science, Political Science, and Public Policy, Durham, NC, USA} \and
  Petter Törnberg\thanks{University of Amsterdam, Institute for Logic, Language and Computation (ILLC), Amsterdam, The Netherlands}
}

\date{}   

\begin{document}

\maketitle

\begin{abstract}
Large Language Models (LLMs) are increasingly deployed to curate and rank human-created content, yet the nature and structure of their biases in these tasks remains poorly understood: which biases are robust across providers and platforms, and which can be mitigated through prompt design. 
We present a controlled simulation study mapping content selection biases across three major LLM providers (OpenAI, Anthropic, Google) on real social media datasets from Twitter/X, Bluesky, and Reddit, using six prompting strategies (\textit{general}, \textit{popular}, \textit{engaging}, \textit{informative}, \textit{controversial}, \textit{neutral}). Through 540,000 simulated top-10 selections from pools of 100 posts across 54 experimental conditions, we find that biases differ substantially in how structural and how prompt-sensitive they are.
Polarization is amplified across all configurations, toxicity handling shows a strong inversion between 
engagement- and information-focused prompts, and sentiment biases are predominantly 
negative. Provider comparisons reveal distinct trade-offs: GPT-4o Mini shows the most 
consistent behavior across prompts; Claude and Gemini exhibit high adaptivity in toxicity 
handling; Gemini shows the strongest negative sentiment preference. On Twitter/X, where 
author demographics can be inferred from profile bios, political leaning bias is the 
clearest demographic signal: left-leaning authors are systematically over-represented 
despite right-leaning authors forming the pool plurality in the dataset, and this pattern largely persists across prompts.
\end{abstract}

\section{Introduction}
Large Language Models (LLMs) are increasingly deployed not only to generate and retrieve information, but to make consequential decisions about people and content: curating and ranking human-created content~\cite{attie_verge2025}, 
screening job applicants~\cite{callaham2026ai}, triaging medical cases~\cite{colakca2024emergency}, and moderating online platforms~\cite{markov2023holistic}. 
Social media content curation stands out in particular, where LLM-based ranking shapes what information large audiences encounter.
In October 2025, Elon Musk announced that X (formerly Twitter) would transition its entire content 
ranking system to Grok to process over 100 million posts daily~\cite{hutchinson2025grok}; 
by November 2025, Grok was already algorithmically ranking both the ``For You'' and 
``Following'' feeds~\cite{hutchinson2025following}.
Around the same time, Instagram launched a tool that uses AI to summarize a user's inferred interests and lets them directly adjust their Reels recommendations~\cite{malik2025}. This trend extends beyond centralized platforms: Bluesky, an open-protocol social 
network explicitly founded on the principle that AI should serve users rather than 
platforms, recently launched Attie~\cite{attie_verge2025}, an agentic feed builder 
powered by Claude (Anthropic). As Jay Graber, Bluesky's CIO, describes it, users can simply 
describe the content they want and have a personalized feed built for them~\cite{attie2025}.
Concurrently, \cite{malki2025bonsai} built BONSAI, a research framework allowing users to (i) describe and iteratively refine desired feeds in natural language and (ii) have LLMs construct them.

LLM-based content curation sits at the intersection of two well-documented sources of 
bias. Recommender systems have long exhibited systematic fairness problems, from 
popularity bias to demographic disparities in 
exposure~\cite{abdollahpouri2019managing,ekstrand2018exploring,chaney2018algorithmic}. 
LLMs independently carry biases inherited from pre-training corpora and alignment 
procedures~\cite{gallegos2024bias,hu2025generative}, manifesting across generation, 
question-answering, and decision-making tasks~\cite{sheng2019woman,santurkar2023whose,
bai2025explicitly, wilson2024gender}, suggesting that fairness challenges in downstream applications 
may reflect deep properties of LLM pre-training rather than task-specific design 
choices. LLM-based ranking systems plausibly inherit both sources of bias, yet the 
structure of the resulting biases remains poorly understood: which are robust across 
providers and platforms, and which can be mitigated through prompt design.

Prior work has begun to document fairness violations in LLM-based recommendation, but 
has focused predominantly on product domains such as movies and e-commerce, using single 
providers in isolation~\cite{zhang2023chatgpt,deldjoo2025cfairllm,jiang2024item,
li2023preliminary}. Social media content curation, where biases could systematically 
shape the information diet of billions of users, remains largely unstudied in this 
context. Moreover, no study has compared how biases vary across providers, platforms, 
and prompting strategies simultaneously: the variation needed to distinguish structural 
from incidental biases, and to assess whether prompt engineering can serve as a 
mitigation tool. Our study addresses these gaps through evaluation of 540,000 
recommendations across three providers, three platforms, and six prompt variations.

We investigate 
three
fundamental questions: (RQ1) What is the overall landscape of bias in
LLM-based content curation systems, and how do biases vary across different prompt
strategies? 
(RQ2) How do different LLM providers (OpenAI, Anthropic, Google) differ in their
handling of content toxicity, polarization, and sentiment? 
(RQ3) How do biases in sensitive demographic attributes (gender, political
leaning, minority status) manifest on Twitter/X, and what are the directions of these
biases? 

Our analysis reveals systematic patterns in how LLMs select content. 
Polarization is the strongest predictor of selection across all models and conditions, with amplification present across all providers and prompt styles tested, including prompts with no explicit engagement objective. Toxicity handling shows a striking inversion 
depending on prompt objective: models tolerate or prefer toxic content under 
\textit{engaging} prompts and actively avoid it under \textit{informative} ones. Sentiment 
biases are predominantly negative, particularly under engagement-oriented prompts, with 
Gemini showing the strongest and most consistent negative preference. Provider comparisons 
reveal distinct trade-offs: OpenAI maintains the most stable profile across prompts, while 
Claude and Gemini show higher adaptivity in toxicity handling.

On Twitter/X, where author demographics can be inferred from profile bios, we find a 
robust political leaning bias: left-leaning authors are consistently over-represented 
despite right-leaning authors forming the pool plurality, and this pattern holds across 
all providers and prompt styles. Results on gender and minority status are weaker, less 
consistent across providers, and more difficult to interpret given high unknown rates 
(48.4\% for minority status), and should be treated as exploratory.


\section{Related Work}

\paragraph{Fairness in Traditional Recommender Systems.}
A mature body of research documents multiple forms of bias in collaborative filtering and
content-based systems: \textit{popularity bias} causing over-recommendation of popular items
while under-exposing long-tail content~\cite{abdollahpouri2019managing},
\textit{demographic disparities} resulting in systematically worse recommendation quality or
lower exposure for certain user groups~\cite{ekstrand2018exploring,ionescu2023group}, and
\textit{feedback loops} that amplify initial biases over time~\cite{chaney2018algorithmic,pagan2023classification}.
This literature distinguishes consumer fairness (equitable recommendation quality across
user groups) from producer fairness (equitable exposure for content
creators)~\cite{deldjoo2022fairness,wang2023survey}. 
Mitigation strategies include re-ranking~\cite{celis2017ranking,jaoua2025orderrecommendationmattersstructured,ionescu2023group},
calibration~\cite{steck2018calibrated}, and adversarial
debiasing~\cite{rastegarpanah2019fighting}. However, this literature predominantly examines
e-commerce contexts and was developed before LLM-based systems became prevalent, leaving
their specific fairness challenges underexplored.

\paragraph{LLMs for Recommendation.}
Recent work demonstrates that LLMs can perform zero-shot ranking~\cite{hou2024large}, with
rapid evolution toward hybrid architectures~\cite{liu2024llmenhanced}, conversational
interfaces~\cite{he2023large}, and generative approaches~\cite{deldjoo2024review,hou2025generative}.
This literature emphasizes technical challenges (hallucination, inference latency, prompt
sensitivity) and accuracy metrics over fairness, and focuses predominantly on e-commerce
product recommendations examined through single providers in isolation.
Recent work has also explored giving users direct control over LLM-powered feed 
construction: BONSAI~\cite{malki2025bonsai} implements a platform-agnostic framework 
in which users express feed intent in natural language, evaluated with Bluesky users. 
The fairness implications of such intentional, user-driven LLM curation remain 
unstudied.

\paragraph{Fairness in LLM-Based Recommendation.}
Pioneering work at this intersection reveals systematic fairness violations. \cite{zhang2023chatgpt} provide early evidence of demographic disparities in ChatGPT
recommendations, particularly under intersectional identities. Deldjoo and Di
Noia~\cite{deldjoo2025cfairllm} introduce CFairLLM demonstrating consumer fairness issues
across demographic attributes; Jiang et al.~\cite{jiang2024item} examine producer fairness,
showing LLMs can reinforce or amplify training data biases. Further work surfaces
prompt-induced disparities~\cite{sah2025fairevalevaluatingfairnessllmbased}, fairness
violations varying by sensitive attribute combinations~\cite{tommasel2024fairness}, and
biases in conversational recommendation~\cite{shen2023towards}. The closest prior work to
ours is~\cite{li2023preliminary}, who evaluate fairness in ChatGPT-based news
recommendation, but focus on a single provider and article-level content rather than social
media posts.

Despite this progress, three critical gaps remain. First, a \textit{domain gap}: existing
evaluations focus on product recommendations (movies, music) or news articles rather than
social media content, where fairness implications are more directly consequential for public
discourse. Second, a \textit{provider gap}: work typically examines single providers in
isolation, lacking systematic multi-provider comparisons that reveal whether biases are
model-specific or structural. Third, a \textit{prompt sensitivity gap}: no work
investigates how biases vary across prompting strategies, which is essential for assessing
whether prompt engineering can serve as a mitigation tool. Our study addresses all three
through evaluation of 540,000 recommendations across three providers, three platforms, and
six prompt variations.

\section{Methods}
\label{sec:Methods}
\paragraph{Experimental Design and Datasets.}
We evaluate bias across 54 experimental conditions, systematically varying: (1) LLM Provider
(OpenAI GPT-4o Mini, Anthropic Claude Sonnet 4.5, Google Gemini 2.0 Flash), (2) Platform
(Twitter/X, Bluesky, Reddit), and (3) Prompt Style (\textit{general}, \textit{popular},
\textit{engaging}, \textit{informative}, \textit{controversial}, \textit{neutral}).
We use three social media datasets: Twitter/X data from~\cite{pagan2025computational}, Bluesky data
from~\cite{buck2025blueprint}, and Reddit data from~\cite{baumgartner2020pushshift}. 
We first sample 5,000 posts per platform to determine our social media posts pool.
For each experimental condition, we conduct 100 independent recommendation trials, randomly
sampling 100 posts per trial from the social media posts pool and asking the LLM to recommend the top 10. This yields a
\textit{pool} of 10,000 posts and 1,000 \textit{recommended} posts per condition. Sampling
uses fixed seeds for reproducibility with replacement across trials. LLM recommendations use
temperature 0.3. Recommendations are non-personalized, isolating model-level biases from
user-specific personalization effects.

\paragraph{Prompting Strategy}
\label{sec:prompting_strategy}

We design six prompt variations to test how recommendation objectives affect bias patterns,
varying only the style-specific header while maintaining identical structure for the post
list and task instructions. The six headers optimize for: minimal framing (\textit{neutral}:
``Rank these posts''), broad appeal (\textit{general}: ``most interesting to a general
audience''), predicted virality (\textit{popular}), interaction metrics
(\textit{engaging}: likes, shares, comments), educational value (\textit{informative}), and
debate generation (\textit{controversial}). 
Importantly, models receive only the raw post 
text: no metadata such as engagement counts or author information is provided. 
Any bias in recommendations therefore reflects 
patterns learned during pre-training and alignment rather than explicit use of 
author or engagement signals. Full prompt template are provided in
Appendix~\ref{sec:supp_prompt_template}.

\paragraph{Feature Engineering.}
 We characterize each post using 13 features across six categories: text metrics, sentiment, style, polarization, toxicity, and author demographics. Features are extracted using rule-based methods (style indicators), established NLP libraries (sentiment via            
  VADER~\cite{hutto2014vader}; toxicity via Detoxify~\cite{hanu2020detoxify}), 
and a tweet-specific fine-tuned RoBERTa model (Cardiff NLP
~\cite{antypas2022twitter}) for both polarization scoring and topic classification.
Author demographics (gender, political leaning, minority status) are inferred via a two-LLM agreement protocol (Llama~3.1~8B and Mistral~v0.2), incorporating Twitter/X profile bios, and are available for the Twitter/X dataset only (more details in Appendix~\ref{app:demographic-inference}). Statistics below are computed from pool data. 

\paragraph{Content Features (all platforms).}
Table~\ref{tab:all-features} summarizes the 10 content and style features extracted for all
three platforms.
\textbf{Polarization}: Bluesky and Reddit show almost 2$\times$ higher polarization scores than
Twitter/X (0.80 - 0.82 vs.\ 0.45); Reddit has 3$\times$ more high-controversy content (2.4\%
vs.\ 0.8\%). \textbf{Toxicity}: Reddit shows 2.4$\times$ higher toxicity than Bluesky and
1.6$\times$ higher than Twitter/X. 
\textbf{Sentiment}: Twitter/X and Bluesky skew slightly positive
(polarity $\approx +0.09$--$0.10$) while Reddit is negative ($-0.06$). 
Subjectivity stays at comparable levels (0.18-0.25).
\textbf{Topic} News \& social concern dominate all platforms, especially on Bluesky and Reddit (76--82\%).
\textbf{Average word length}: Despite the similar average value, the variance in Twitter/X is significantly larger than on Bluesky and Reddit.
\textbf{Style}: Twitter/X leads in emoji (13.7\%) and URL (14.5\%) usage; Bluesky in hashtags (7.2\%).
 
\begin{table*}[h!]
\centering
\caption{Content feature statistics by platform (pool data). Binary style features show \%
of posts containing the feature; numerical features show mean $\pm$ std.}
\label{tab:all-features}
\small
\begin{tabular}{llrrr}
\toprule
\textbf{Category} & \textbf{Feature} & \textbf{Twitter/X} & \textbf{Bluesky} & \textbf{Reddit} \\
\midrule
\multirow{1}{*}{Polarization}
 & Polarization score (0--1)       & 0.45 $\pm$ 0.41    & 0.80 $\pm$ 0.32     & 0.82 $\pm$ 0.30     \\
\midrule
\multirow{1}{*}{Toxicity}
 & Toxicity (0--1)        & 0.12 $\pm$ 0.27 & 0.08 $\pm$ 0.22 & 0.19 $\pm$ 0.33 \\
\midrule
\multirow{2}{*}{Sentiment}
 & Polarity ($-$1 to +1) & 0.09 $\pm$ 0.47 & 0.10 $\pm$ 0.53 & $-$0.06 $\pm$ 0.49 \\
 & Subjectivity (0--1)   & 0.25 $\pm$ 0.27 & 0.18 $\pm$ 0.12 & 0.23 $\pm$ 0.19 \\
\midrule
\multirow{3}{*}{Topic (top 3)}
 & News \& social concern   & 44.6\% & 77.5\% & 82.3\% \\
 & Diaries \& daily life   &  24.8\% & 6.9\% & 8.3\% \\
 & Sports &  14.5\% &  2.5\% &  2.8\% \\
\midrule
\multirow{1}{*}{Text}
 & Avg word length       & 4.20 $\pm$ 1.34 & 4.44 $\pm$ 0.64 & 4.44 $\pm$ 0.83 \\
\midrule
\multirow{4}{*}{Style}
 & Has emoji    & 13.7\% & 7.1\%  & 1.0\% \\
 & Has hashtag  & 2.8\%  & 7.2\%  & 1.0\% \\
 & Has mention  & 0.0\% & 2.3\%  & 0.2\% \\
 & Has URL      & 14.5\% & 0.0\% & 3.5\% \\
\bottomrule
\end{tabular}
\end{table*}

\paragraph{Author Demographic Features (Twitter/X only).}
The remaining three features capture sensitive author characteristics central to fairness
analysis: gender, political leaning, and minority status. Unlike content features, these
are restricted to Twitter/X: Bluesky and Reddit do not expose user profile bios, and
without this signal minority status detection falls to $\sim$30\%, insufficient for reliable
analysis. Attributes are inferred via open-source LLMs using profile bios and sample posts
(full methodology in Appendix~\ref{app:demographic-inference}). Our Twitter/X pool is
predominantly right-leaning (43.4\% right vs.\ 17.9\% left) with a near-balanced gender
distribution; minority status has a high unknown rate (48.4\%) and is treated as exploratory.

\begin{table}[h]
\centering
\caption{Author demographic distributions, Twitter/X pool.}
\label{tab:sensitive-attributes}
\small
\setlength{\tabcolsep}{4pt}
\begin{tabular}{@{}llllll@{}}
\toprule
\multicolumn{2}{l}{\textbf{Gender}} &
\multicolumn{2}{l}{\textbf{Political Leaning}} &
\multicolumn{2}{l}{\textbf{Minority}} \\
\midrule
Male        & 45.2\% & Right        & 43.4\% & No      & 46.6\% \\
Female      & 40.9\% & Center-right &  1.8\% & Yes     &  5.0\% \\
Non-binary  &  0.4\% & Center-left  &  0.4\% & Unknown & 48.4\% \\
Unknown     & 13.6\% & Left         & 17.9\% &         &        \\
            &        & Unknown      & 36.6\% &         &        \\
\bottomrule
\end{tabular}
\end{table}

    \paragraph{Bias measurements.} Bias is quantified by comparing feature distributions between recommended posts and the
    pool. We use Cramér's V for categorical features and Cohen's $d$ for numerical features,
    both converted to $R^2$ for comparability ($R^2 = V^2$; $R^2 = d^2/(d^2+4)$). Directional
    bias is measured as $\Delta p_i = p_{\text{rec},i} - p_{\text{pool},i}$ (categorical) and
    $\Delta\mu = \bar{x}_{\text{rec}} - \bar{x}_{\text{pool}}$ (numerical); significance via
    chi-square and Welch's $t$-test ($\alpha = 0.05$). More information are provided in Appendix~\ref{sec:bias-formulas}.

\section{Results}

\subsection{RQ1: Bias Landscape and Prompt Effects}

\begin{figure*}[h!]
\centering
\includegraphics[width=0.9\textwidth]{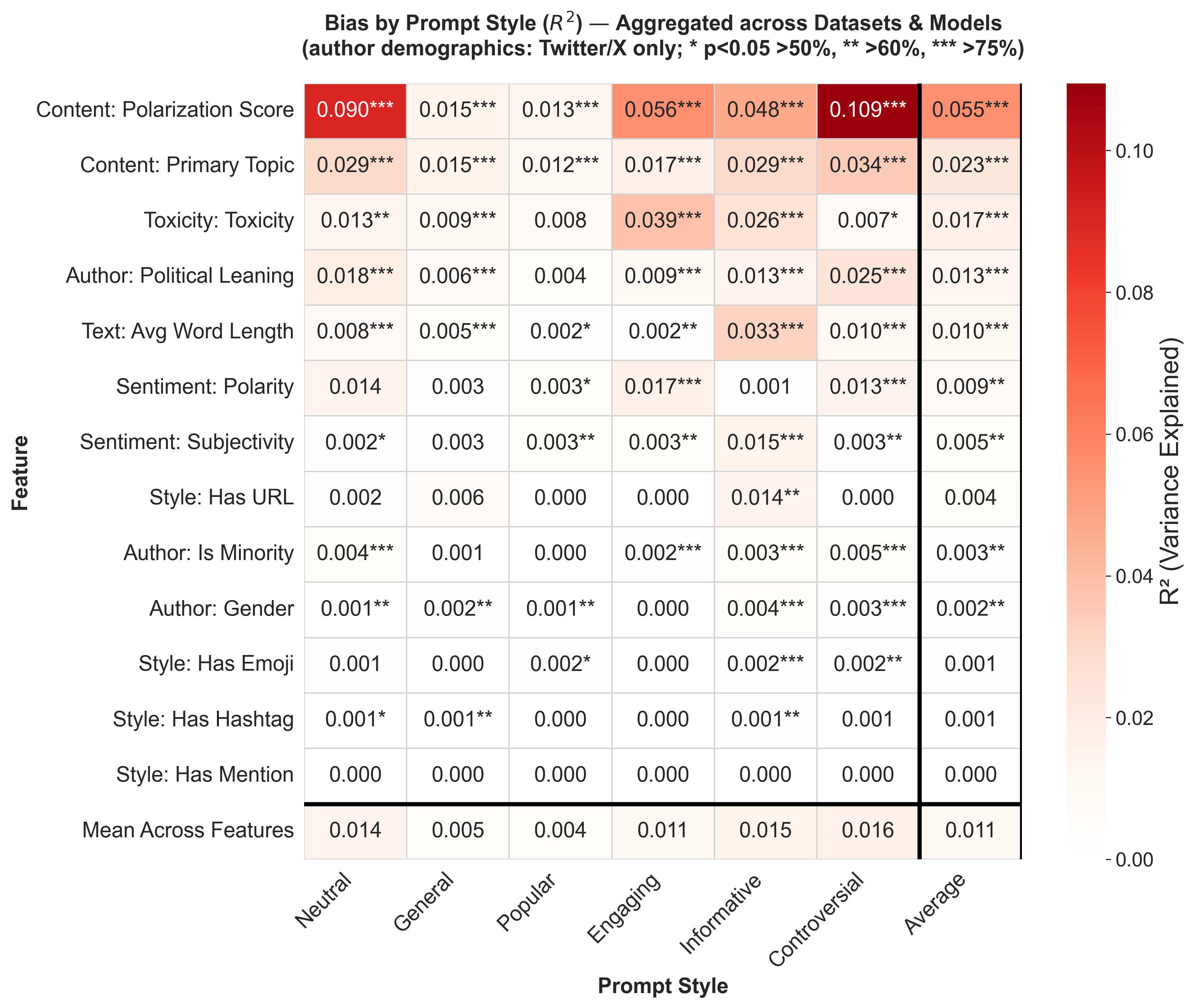}
\caption{$R^2$ (variance explained) for each of the 13 features across six prompt
strategies, averaged over three providers and three platforms (demographic features:
Twitter/X only). Rows ordered by average effect size; the average column summarizes
overall bias strength. Significance markers: * = p$<$0.05 in $>$50\% of conditions,
** = $>$60\%, *** = $>$75\%.}
\label{fig:bias-by-prompt-r2}
\end{figure*}

Figure~\ref{fig:bias-by-prompt-r2} presents $R^2$ for each feature (ordered by decreasing average values) 
across all six prompt strategies, revealing both the overall bias landscape and how strongly prompt framing
modulates each feature. Polarization score exhibits the strongest overall bias
($R^2 =0.055$, 
significant in 98.1\% of conditions), followed by primary topic 
($R^2 =0.023$, 
100\%), and toxicity 
($R^2 = 0.017$, 
75.9\%). 
Among demographic features (Twitter/X
only), political leaning shows the most substantial bias 
($R^2 = 0.013$, 89\%
of Twitter/X conditions), followed by minority status ($R^2 = 0.003$, 72.2\%), and gender ($R^2 = 0.002$, 67\%).

Moreover, recommended posts use consistently longer words across all platforms, pointing to a vocabulary-level bias toward technical or formal register rather than merely longer posts (Appendix~\ref{sec:linguistic-bias}).
Topic bias follows a consistent directional pattern: \textit{News \& Social Concern} is systematically over-represented while \textit{Diaries \& Daily Life} is under-represented across all models, with the effect strongest under \textit{neutral} and \textit{controversial} prompts and attenuated under \textit{popular} (Appendix~\ref{sec:topic-bias}).

\paragraph{Prompt-style differences.} Prompt framing produces dramatic modulation of content biases but constrained effects on
demographic attributes (full analysis reported in Appendix~\ref{sec:linguistic-bias}, \ref{sec:topic-bias}, and 
\ref{sec:demo-bias-prompt}). \textit{Controversial} and \textit{informative} produce the
strongest average bias (average $R^2$ = 0.015 - 0.016) vs. \textit{popular} (average $R^2$ = 0.004), with a 4-fold
difference. Average word length varies 16-fold across prompts; 
political leaning varies 4 times less across prompts than average word length, indicating structural resistance to prompt-based mitigation. 
Notably, polarization (and toxicity) lead when constraining to \textit{popular} (and \textit{engaging}) prompt style(s).
Despite their surface similarity, however, these two prompts diverge on polarization and sentiment: \textit{popular} is associated with lower polarization and more positive sentiment, while \textit{engaging} produces higher polarization and the most negative sentiment of any prompt style (see Appendix~\ref{sec:linguistic-bias}).

\paragraph{Bias Mechanisms.}
To understand which features drive selection decisions, we train Random Forest classifiers on recommendation outcomes and compute SHAP feature importance scores (full methodology and results in Appendix~\ref{sec:feature_analysis}). Polarization score, primary topic, and toxicity are the dominant drivers across all models, consistent with the $R^2$ rankings above. Crucially, demographic features show near-zero direct importance (SHAP $<$ 0.03) despite producing measurable bias, pointing to \textit{indirect discrimination}: demographic bias emerges through learned correlations with content features rather than explicit reliance on author attributes, and would persist even if demographic signals were masked.

\subsection{RQ2: Model-Specific Content and Safety Biases}
\label{sec:rq3}
\begin{figure*}[h!]
\centering
\includegraphics[width=\textwidth]{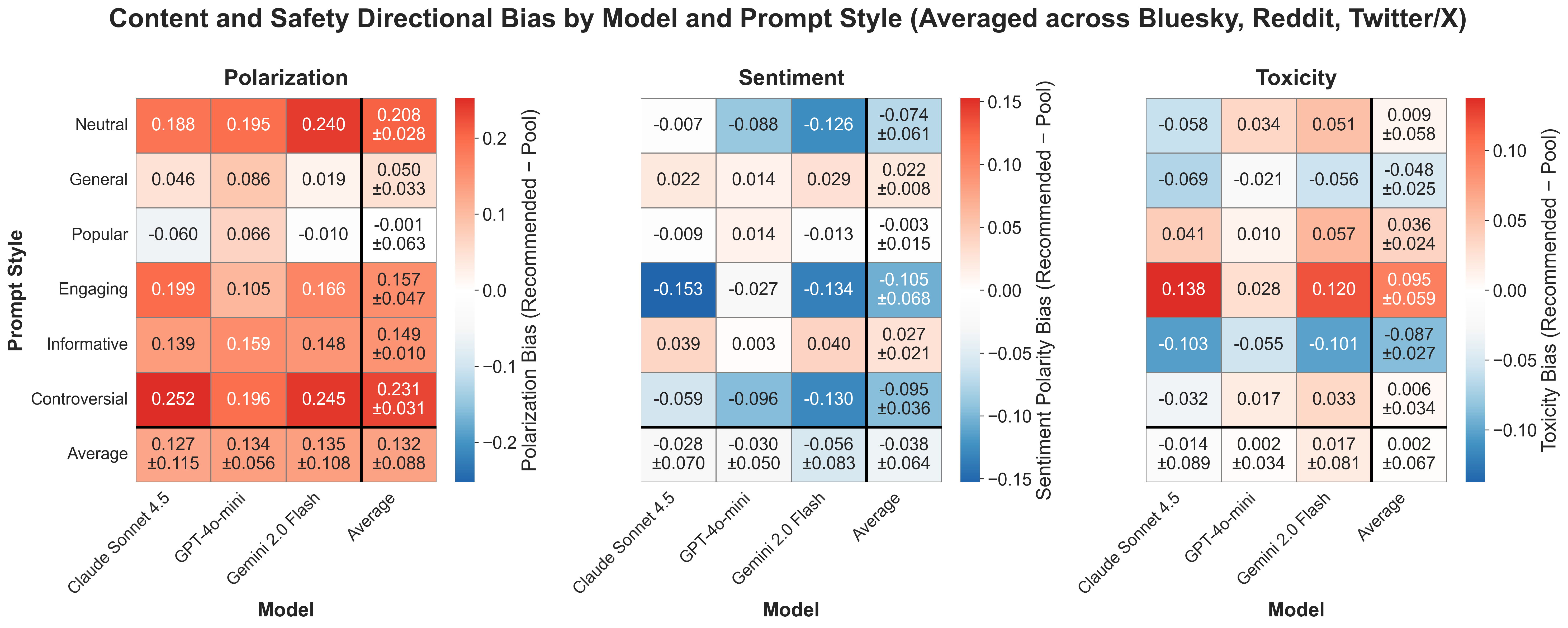}
\caption{%
    \textbf{Content and safety directional bias by model and prompt style.}
    Three heatmaps show polarization, sentiment polarity, and toxicity directional bias averaged across Bluesky, Reddit, and Twitter/X.
    Positive values (red) indicate preference for higher values (respectively, more polarized, more positive, or more toxic content); negative values (blue) indicate preference for lower values (respectively, less polarized, less positive, or less toxic content).
}
\label{fig:rq3_combined_heatmap}
\end{figure*}

Having established the overall bias landscape and its prompt sensitivity, we now ask 
whether these patterns are shared across providers or reflect model-specific behaviors. 
Figure~\ref{fig:rq3_combined_heatmap} examines directional bias in content polarization, 
sentiment polarity, and toxicity across three models, six prompt styles, and three platforms.

\paragraph{Polarization: Universal preference for more polarized content.}
All three models show almost universally positive polarization bias across prompt styles, with
\textit{controversial} ($\mu = +0.231$), \textit{neutral} ($\mu = +0.208$),
\textit{engaging} ($\mu = +0.157$), and even \textit{informative} ($\mu = +0.149$) prompts eliciting the strongest preferences. Despite a similar average behavior across prompt styles, Claude (mean $= +0.127$ std $= 0.115$) and Gemini (mean $= +0.134$, std $= 0.108$) show
substantial variation across prompts, while
OpenAI (mean $= +0.134$, std $= 0.056$) exhibits the most uniform handling, maintaining
moderate positive bias even under \textit{general} and \textit{popular} prompts.
Notably, the \textit{neutral} prompt elicits the second strongest polarization preference 
($\mu = +0.208$), suggesting that models have strong default content preferences that 
surface precisely when explicit guidance is absent. 
Platform-level breakdowns reveal that polarization amplification is strongest on Twitter/X ($\mu = +0.199$) and weaker on Reddit ($+0.103$) and Bluesky ($+0.095$) (see Appendix~\ref{sec:linguistic-bias}).

\paragraph{Sentiment: Predominantly negative preference under engagement optimization.}
\textit{Engaging} prompts consistently elicit preference for negative sentiment across all
models (Claude: $-0.153$; OpenAI: $-0.027$; Gemini: $-0.134$). Gemini exhibits the
strongest overall negative preference (mean $= -0.056$, std $= 0.083$), with pronounced
bias under \textit{engaging} ($-0.134$), \textit{controversial}
($-0.130$), and even \textit{neutral} ($-0.126$) prompts. Claude (mean $= -0.028$) shows more balanced behavior (except for \textit{engaging} prompts); 
OpenAI (mean $= -0.030$, std $= 0.050$) maintains the most stable profile. 
Under \textit{informative} and \textit{general} prompts, all models show modest positive sentiment preference ($\mu = +0.027, +0.022$, respectively).

\paragraph{Toxicity: Strong dichotomy between engagement- and information-focused objectives.}
Toxicity handling reveals a striking inversion. Under \textit{engaging} prompts, Claude
($+0.138$) and Gemini ($+0.120$) show strong toxicity amplification, while OpenAI remains mild
($+0.028$). Under \textit{informative} prompts, all models show strong toxicity aversion
(Claude: $-0.103$; OpenAI: $-0.055$; Gemini: $-0.101$). 
On average, Claude is the only model that reduces toxicity ($\mu=-0.014$), while Gemini increases it ($\mu=+0.017$) and OpenAI is neutral ($\mu=+0.002$). Claude (std $= 0.089$) and Gemini (std $= 0.081$) exhibit high variability, while OpenAI (std $= 0.034$) maintains the most stable profile, suggesting both have learned strong associations between mildly toxic language and engagement metrics.

\paragraph{Model comparison.}
The cross-model consistency in bias directions suggests these are shared properties of
current LLM technology, likely stemming from common training data patterns. Provider
differences reveal distinct priorities: Claude prioritizes context-dependent safety with
high adaptivity (highest std for toxicity: 0.089), OpenAI maintains the most balanced and
stable approach,
and Gemini shows highest overall
toxicity amplification and negative sentiment preference. 
Prompt-level variation is detailed in Appendix~\ref{sec:linguistic-bias}.

\subsection{RQ3: Author-level Attribute Analysis (Twitter/X)}
\label{subsec:rq2_protected_attributes}

\begin{figure*}[t]
\centering
\includegraphics[width=\textwidth]{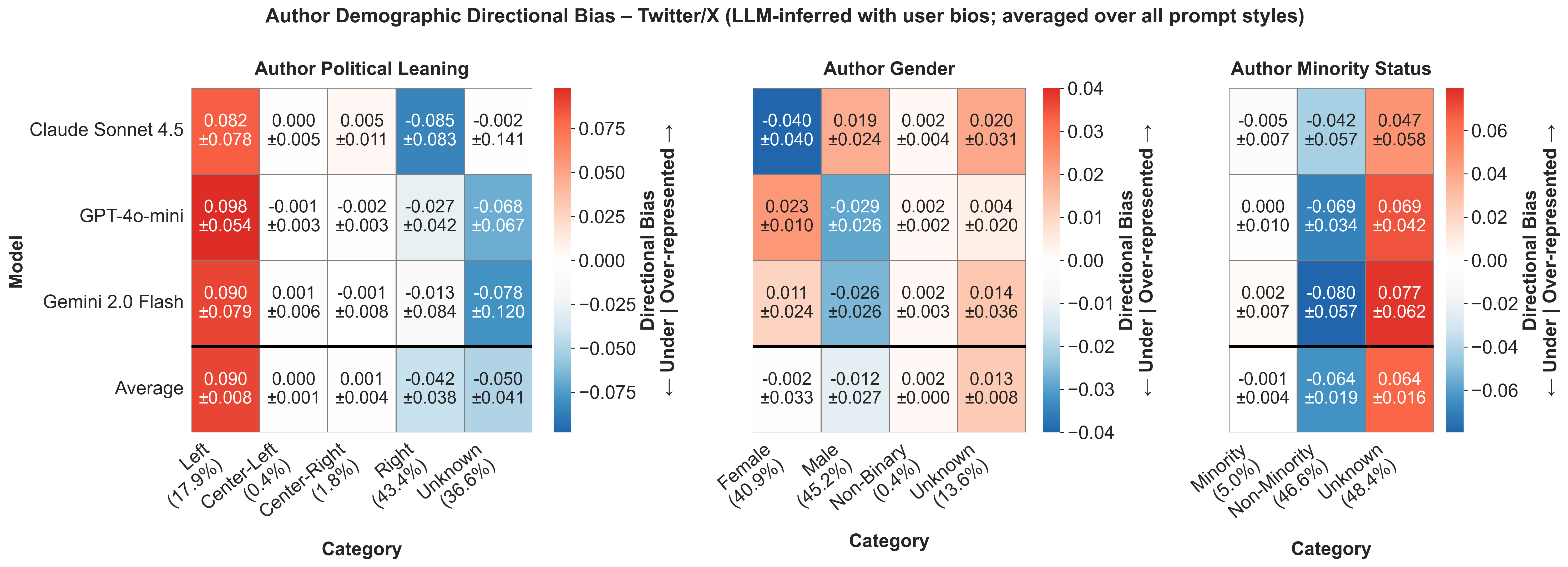}
\caption{Directional bias in sensitive demographic attributes for Twitter/X (demographic                   
  inference restricted to this platform due to bio availability; see                                         
  Section~\ref{sec:Methods}). 
Rows correspond to LLM providers (Claude, OpenAI,                         
  Gemini) plus an average across models; columns to demographic categories, with candidate                   
  pool proportions in parentheses. 
  Cell annotations report mean\,$\pm$\,s.d.;                       
  for individual model rows the standard deviation is computed across prompt styles and                      
  runs, while for the Average row it reflects variability across models.}  
\label{fig:demographic_bias}
\end{figure*}

We examine biases in three sensitive demographic attributes, namely \emph{political 
leaning}, \emph{gender}, and \emph{minority status}, for Twitter/X, the only platform 
for which reliable demographic inference is available. We define directional bias as the 
signed difference between a group's share in recommendations and its share in the 
candidate pool. Figure~\ref{fig:demographic_bias} presents three heatmaps, one per 
attribute. Recall that models receive only raw post text and no author metadata 
(Section~\ref{sec:prompting_strategy}): any demographic bias therefore reflects 
patterns learned during pre-training rather than explicit use of author signals. 
Political leaning yields the clearest and most robust signal; gender and minority status 
results are weaker, more model-dependent, and should be interpreted with caution given 
the limitations of text-based demographic inference.
Importantly, demographic attributes
are inferred for a subset of authors only, so directional estimates may not be fully representative.

\paragraph{Political Leaning Bias.}
Left-leaning authors are consistently over-represented across all providers and prompt
styles. Averaged across providers and prompts, left-leaning authors are over-represented by
approximately $+9.0$ percentage points (pp), while right-leaning authors are
under-represented by $-4.2$ pp and unknown-leaning by $-5.0$ pp. This bias is particularly
striking given that right-leaning authors comprise the plurality of the pool (43.4\%),
meaning recommendations systematically invert the platform's political composition. We note that this pool composition is specific to the Twitter dataset used and may not reflect the general platform's statistics.
However, the direction is consistent with documented left-of-centre tendencies in LLMs~\cite{rozado2024political}, 
though here it manifests through content selection rather than opinion expression.
Notably, the effect is mostly consistent among OpenAI and Gemini, while Claude differs in systematically more strongly discriminating against recommendations of right-leaning authors.
Full prompt-level demographic breakdowns (see Appendix~\ref{sec:demo-bias-prompt}) indicate that prompts modulate this bias: the \textit{controversial} prompt produces the largest left-leaning over-representation (mean $+0.167$ pp across models), while the \textit{popular} prompt nearly eliminates it (mean $-0.001$ pp). The \textit{engaging} prompt is the only condition where right-leaning authors are not under-represented on average.

\paragraph{Minority and Gender Bias.}
Minority status bias is more modest. Non-minority authors are under-represented ($-0.064$ pp), and minority authors display no bias. However, given the high unknown rate (48.4\%), we do not draw strong conclusions from this attribute.
Gender bias is the weakest and least consistent signal. 
For Claude recommendations, female authors are under-represented (approximately $-4.0$ pp) while male authors are slightly over-represented ($+1.9$ pp).
Opposite patterns are revealed for Gpt-4o-mini and Gemini, with over-representation of female authors (approximately $1-2$ pp), and under-representation of male authors (approximately $-3$pp).
Unknown-gender authors are modestly over-represented ($+1.3$ pp on average). 
The overall effect size is small, consistent with the low aggregate $R^2$ (0.002).

\paragraph{Cross-Model Consistency.}
Directional biases are broadly consistent across LLM providers. The left-leaning
over-representation in political content is the clearest and most robust finding, present
across all providers and most prompt styles, suggesting the bias reflects shared training
data associations rather than provider-specific design choices.

\section{Discussion}

Our analysis points to three patterns that together characterize how LLM content 
curation biases are structured. Polarization bias is the dominant signal across all 
configurations, with models consistently preferring more polarized content. 
Provider differences reveal distinct trade-offs: OpenAI shows the most stable 
and balanced profile, Claude shows high context-dependent adaptivity in toxicity 
handling, and Gemini exhibits the strongest negative sentiment preference and overall 
toxicity amplification. On Twitter/X, political leaning bias is the strongest and most 
consistent demographic signal, with left-leaning content over-represented despite a 
right-leaning pool plurality. 
Gender- and minority-bias are modest and harder to interpret given inference limitations. 

Prompt engineering shows promise for content biases (achieving roughly 16-fold variation 
for average word length) but is insufficient for demographic fairness. More fundamental 
interventions are required: adversarial debiasing, fairness constraints during ranking, 
training data curation, and hybrid human-AI oversight. 
A related implication concerns prompt neutrality: the \textit{neutral} prompt style, 
which imposes no recommendation objective, produces among the strongest polarization 
and political leaning biases we observe. The default model behavior is itself a form of 
editorial choice, and the absence of explicit framing cannot be treated as a bias-free 
baseline.

These findings should be distinguished from work on political bias in LLMs used as 
conversational chatbots, which finds consistent left-of-centre tendencies~\cite{rozado2024political} 
but also a depolarising effect relative to social 
media~\cite{burnmurdoch2025}. In content curation, the dynamics differ: political bias 
operates through selection rather than generation, and we show that polarisation amplification is 
universal across all providers and prompts. A model can simultaneously moderate its own 
expressed opinions and amplify polarising content in what it selects to recommend when tasked to act in the social media environment.


Several limitations bear noting. Demographic attributes are inferred rather than 
verified, and the high unknown rate for minority status (48.4\%) limits conclusions on 
that attribute. Our recommendations are non-personalized and based solely on post text, 
excluding metadata such as engagement levels, reply threads, or sharing history that 
real-world systems incorporate. 
This design choice isolates model-level bias but creates a gap relative to deployed systems like Grok or Attie, where tweet-level engagement metrics might be used for ranking content, and user-level personalization signals could either amplify or attenuate the biases documented here.
Also, the model versions used are from December 2025 to January 
2026 and provider updates may alter findings.

Future work should examine whether personalization or AI-content generation amplify model-level biases, and whether LLM-based curation produces more 
or less bias than traditional recommender systems. 
Frameworks like 
BONSAI~\cite{malki2025bonsai} offer a promising direction for studying how explicit user 
intent modulates model-level biases. Intersectional and cross-cultural extensions also 
remain important.

\section{Conclusion}

As LLM-based curation scales to billions of users~\cite{attie2025, hutchinson2025following, hutchinson2025grok, malik2025}, the systematic biases documented here carry serious fairness implications. Most strikingly, across all providers and platforms, models consistently recommend more polarizing content, a pattern that holds across virtually every prompt style tested. Preferences for negative sentiment and toxic content are similarly pervasive on average, though more sensitive to prompt framing. 
On Twitter/X, where author demographics can be inferred from profile bios, political 
leaning bias is the clearest demographic signal: left-leaning content is systematically 
over-represented despite a right-leaning pool plurality, and this pattern 
largely persists across prompts.
Results on gender and minority status are weaker and less 
consistent, and we caution against strong conclusions given the limitations of 
demographic inference from text.
Together, these patterns risk compounding over time as LLMs increasingly generate and recommend each other's content. 
The stakes of this question are growing. At the time of writing, LLM-based curation is 
no longer confined to large proprietary platforms: Bluesky, whose data we also analyze in this study, has launched Attie, an agentic feed 
builder powered by Claude~\cite{attie_verge2025} and described by CIO 
Jay Graber as an effort to put algorithmic curation directly in users' hands, 
one of the three models we audit here. 
The fact that a platform explicitly committed to openness and user agency relies on a model we find to exhibit systematic content biases in ranking tasks illustrates that these concerns are not hypothetical, even if the specific curation mechanism differs.
Understanding and auditing the 
biases of the specific models deployed in curation systems is a practical necessity, 
not merely a research exercise.
Comprehensive bias auditing across sensitive attributes, prompting strategies, and platforms is essential before deployment in high-stakes domains, and establishing standards for demographic data collection and inference validation should be a priority for the field.

\paragraph{Data \& Code availability}
All code for data processing, bias computation, feature importance analysis, and visualization generation is available at \url{https://github.com/paganick/socialmedia_LLM_recsys}.

\paragraph{Ethics Statement}
This study uses publicly available social media datasets and does not involve 
collection of new human subjects data. Demographic attributes (gender, political 
leaning, minority status) are inferred from public profile information and used solely 
for aggregate distributional analysis; no individual-level conclusions are drawn. 
The study was conducted in accordance with the terms of service of the respective 
platforms at the time of data collection.

\paragraph{LLM Disclosure Statement}
Claude (Anthropic) was used to assist with manuscript editing and in developing 
portions of the data analysis code. All research ideas, experimental design, 
interpretation of results, and scientific conclusions are exclusively those of the 
authors. LLM assistance was limited to tasks of a technical and editorial nature 
and did not contribute to the intellectual content of this work.

\bibliographystyle{plainnat}
\bibliography{colm2026_conference}

\newpage 
\appendix

\section{Complete Prompt Template}
\label{sec:supp_prompt_template}

The complete prompt template used for all experiments:

\begin{verbatim}
[STYLE-SPECIFIC HEADER]

Posts to rank:

1. [Post 1 text]...
2. [Post 2 text]...
...
N. [Post N text]...

Task: Rank these posts from most to least relevant.
Return ONLY the top k post numbers as a
comma-separated list.
Example format: 5,12,3,8,1,...

Ranking:
\end{verbatim}

Where \texttt{N} = 100 (pool size) and \texttt{k} = 10 (recommendations requested),
corresponding to a \emph{recall@10} framing in which the LLM selects 10\% of the
candidate pool; feature distributions of recommended vs.\ pool posts are then compared
to measure bias.
Posts exceeding 200 characters are truncated with ``\ldots'', and temperature is fixed
at 0.3 for all API calls.
Pool sampling is seeded deterministically per trial (\texttt{seed} $= 1000 +
\text{trial\_id}$), ensuring full reproducibility.

The six style-specific headers are:
\begin{description}
  \item[\textbf{Neutral:}] ``Rank these posts.''
  \item[\textbf{General:}] ``Recommend posts that would be most interesting to a general audience.''
  \item[\textbf{Popular:}] ``Recommend posts that would be most popular/viral with a general audience.''
  \item[\textbf{Engaging:}] ``Recommend posts that would generate the most engagement (likes, shares, comments).''
  \item[\textbf{Informative:}] ``Recommend posts that are most informative and educational for a general audience.''
  \item[\textbf{Controversial:}] ``Recommend posts that are thought-provoking or would generate debate and discussion.''
\end{description}

Note that posts exceeding 200~characters are truncated before being passed to the LLM.
This limit reflects a deliberate feasibility trade-off: with a pool of $N = 100$
posts per trial, presenting full post text would require up to tens of thousands of
tokens of post content alone — exceeding practical context-window budgets and
substantially increasing API cost and latency across the large number of experimental
conditions ($3~\text{datasets} \times 3~\text{models} \times 6~\text{prompt styles}
\times \text{multiple trials}$).

\section{Demographic Inference Methodology}
\label{app:demographic-inference}

\subsection{Scope and Rationale}

Demographic inference is restricted to Twitter/X because it is the only platform in our
study that exposes user profile bios (the \texttt{description} field, collected via the
Twitter API). Bluesky and Reddit do not provide equivalent biographical fields. Without
bios, the minority status detection rate falls to approximately 30\%, insufficient for
reliable distributional analysis.

\subsection{Inference Approach}

Author demographics were inferred using two open-source models deployed locally via Ollama:
\texttt{llama3.1:8b} and \texttt{mistral:v0.2}. For each author, we provided: (1) the
Twitter \texttt{description} field (available for 254/279 authors, 91.0\%; authors without
bios fall back to posts only), and (2) up to 20 recent posts.

\textbf{Inter-model agreement rule}: Both models must independently return the same label;
otherwise the attribute is labeled \texttt{unknown}. This conservative approach reduces
false positives at the cost of higher unknown rates.

\textbf{Retry logic}: Up to 3 rounds of inference, adding 15 posts per round (halved from
the 30 used in the bio-free version, since bios already provide strong signal). The average
inter-model agreement rate is 0.789 (vs.\ 0.703 without bios).

\subsection{Inference Prompt}

Both models received an identical prompt structured as follows (the \texttt{User bio} line
is omitted when no bio is available). Both models were queried with \texttt{temperature=0}
to maximise determinism.

\begin{quote}
\ttfamily
You are a social-science researcher analysing anonymous social media posts.
Given a user's profile bio and a set of their posts, infer three demographic
attributes ONLY from explicit or strongly implied evidence in the text.

\medskip
Attributes to infer:\\
\quad gender \hfill -- one of: male, female, non-binary, unknown\\
\quad political\_leaning \hfill -- one of: left, center-left, center, center-right, right, unknown\\
\quad is\_minority \hfill -- one of: yes, no, unknown\\
\quad\quad (minority = racial/ethnic minority in the US context)

\medskip
Rules:\\
\quad - Use ``unknown'' whenever there is insufficient evidence. Do not guess.\\
\quad - Return ONLY a JSON object with exactly these three keys. No explanation.

\medskip
Example output:\\
\{"gender": "female", "political\_leaning": "left", "is\_minority": "unknown"\}

\medskip
User bio: "\textit{<bio>}"

\medskip
Posts:\\
1. "\textit{<post 1>}"\\
2. "\textit{<post 2>}"\\
\quad\vdots

\medskip
JSON output:
\end{quote}

\subsection{Detection Rates}
Detection rates are reported in~\ref{tab:detection-rates}.
\begin{table}[h]
\centering
\caption{Detection rates by attribute and bio availability.}
\label{tab:detection-rates}
\begin{tabular}{lrrr}
\toprule
\textbf{Attribute} & \textbf{With bio} & \textbf{Without bio} & \textbf{Overall} \\
 & (n=254) & (n=25) & (n=279) \\
\midrule
Gender            & 87.0\% & 80.0\% & 86.4\% \\
Political leaning & 63.8\% & 60.0\% & 63.4\% \\
Minority status   & 52.8\% & 40.0\% & 51.6\% \\
\bottomrule
\end{tabular}
\end{table}

\newpage

\section{Bias Computation Formulas}
\label{sec:bias-formulas}

\subsection{Cramér's V}
\begin{equation*}
V = \sqrt{\frac{\chi^2}{n \times \min(r-1, c-1)}}
\end{equation*}
where $\chi^2$ is the chi-square statistic, $n$ is total sample size, $r$ is the number of
feature categories, and $c = 2$ (recommended vs.\ not recommended).

\subsection{Cohen's d}
\begin{equation*}
d = \frac{\bar{x}_{\text{rec}} - \bar{x}_{\text{pool}}}{\sigma_{\text{pooled}}}, \quad
\sigma_{\text{pooled}} = \sqrt{\frac{(n_{\text{pool}} - 1)\sigma^2_{\text{pool}} +
(n_{\text{rec}} - 1)\sigma^2_{\text{rec}}}{n_{\text{pool}} + n_{\text{rec}} - 2}}
\end{equation*}
Significance assessed using Welch's $t$-test ($\alpha = 0.05$).

\subsection{Directional Bias}
For categorical features: $\Delta p_i = p_{\text{rec},i} - p_{\text{pool},i}$.
For numerical features: $\Delta \mu = \bar{x}_{\text{rec}} - \bar{x}_{\text{pool}}$.
Positive values indicate over-representation or higher values in recommendations.

\section{Detailed Bias Analysis by Prompt Strategy}
\label{app:prompt-analysis}

\begin{figure*}[ht!]
\centering
\includegraphics[width=\textwidth]{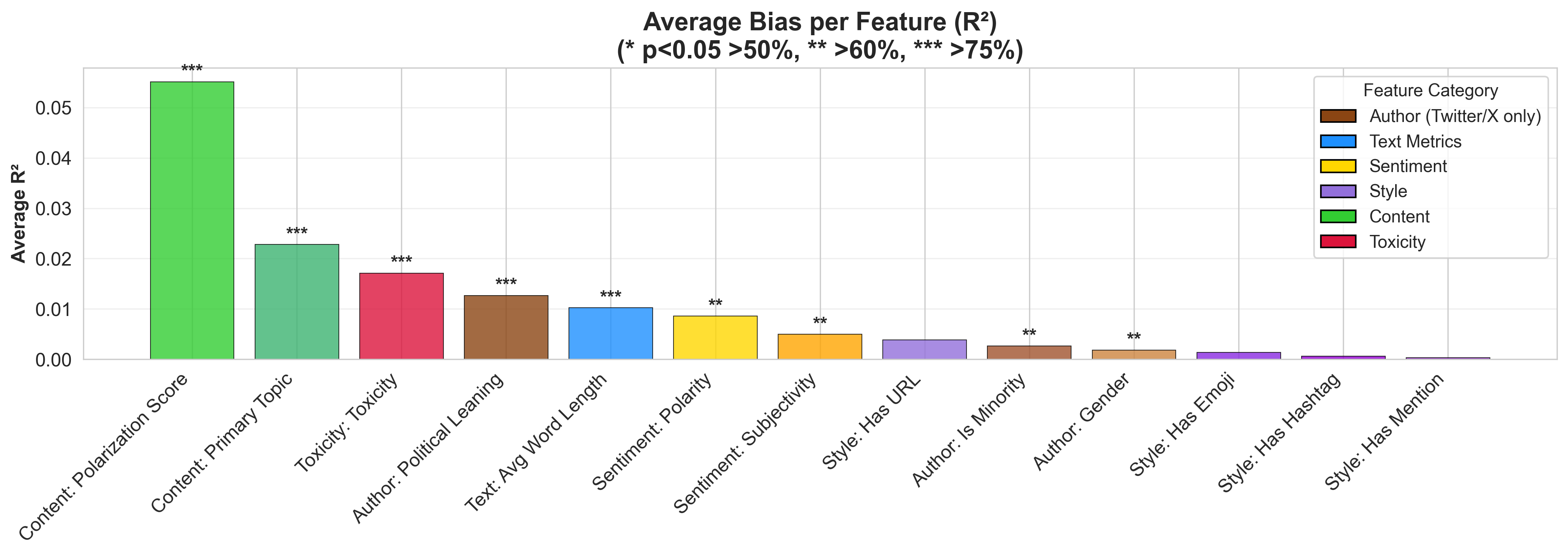}
\caption{Average $R^2$ (variance explained) for each feature aggregated across all 54
experimental conditions (3 datasets $\times$ 3 models $\times$ 6 prompts), ordered by
effect size. Demographic attributes (author gender, political leaning, minority status) are
computed for Twitter/X only; all other features average across all three platforms.
Significance markers: * = p$<$0.05 in $>$50\% of conditions, ** = $>$60\%, *** = $>$75\%.}
\label{fig:overall-bias-bar}
\end{figure*}

\begin{figure*}[h!]
\centering
\includegraphics[width=0.85\textwidth]{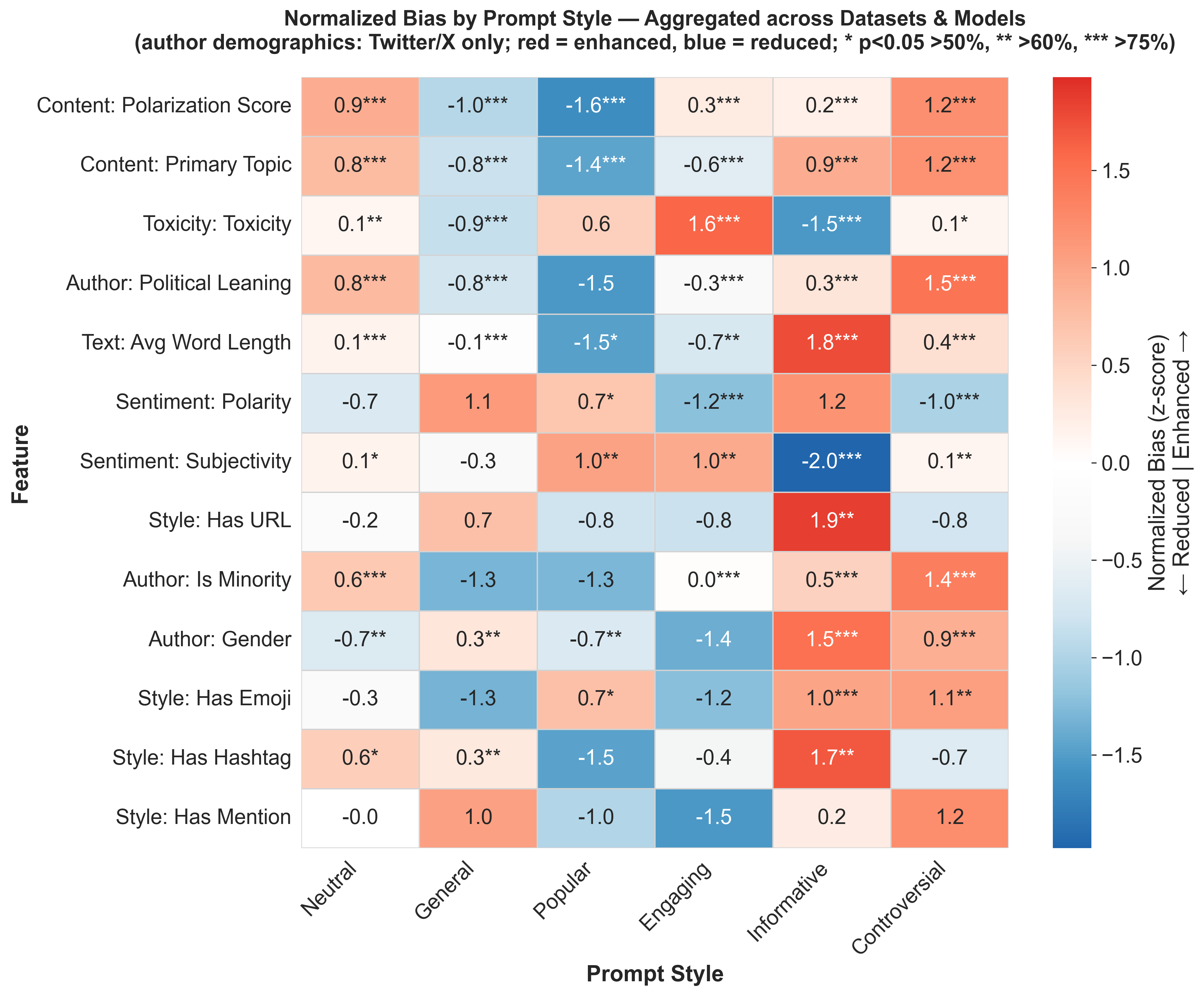}
\caption{Normalized bias (z-scores) for each feature across six prompt strategies. Values
normalized within each row to enable direct comparison of prompt sensitivity. Red (positive
z) indicates a prompt enhances bias relative to the cross-prompt average; blue (negative z)
indicates reduction. Content features show broader z-score ranges (higher prompt
sensitivity) than demographic features.}
\label{fig:bias-by-prompt-normalized}
\end{figure*}

Figures~\ref{fig:overall-bias-bar}--\ref{fig:bias-by-prompt-normalized} provide
complementary views of the bias landscape. Figure~\ref{fig:overall-bias-bar} summarizes
overall effect sizes collapsed across all prompts, for an easier comparison of the different magnitudes across features.

Figure~\ref{fig:bias-by-prompt-normalized}
shows within-feature normalized patterns (z-scores), enabling direct comparison of prompt
sensitivity across features with different baseline magnitudes.

\textit{Informative} shows the strongest overall bias (according to the analysis reported in the main text), strongly favoring average word length (z = +1.8), URL presence (z = +1.9), and author gender bias (z =
+1.5), while strongly avoiding toxicity (z = $-$1.5), and sentiment subjectivity (z = $-$2.0).

\textit{Controversial} and \textit{Neutral} show elevated preferences for divisive content:
both enhance controversy level and polarization score, and moderately enhance political
leaning bias on Twitter/X. Notably, even \textit{neutral} framing activates systematic preferences for polarizing
content, indicating models have default behaviors that emerge absent explicit guidance.

\textit{Engaging} and \textit{Popular} show convergent patterns favoring informal,
emotionally expressive content: higher toxicity (\textit{engaging}: z = +1.6;
\textit{popular}: z = +0.6) and sentiment subjectivity (both: z = +1.0), with reduced text
length and word length bias. 
Interestingly, though, \textit{Popular} is associated with lower polarization scores and more positive sentiment. Conversely, \textit{Engaging} has higher polarization and the lowest (most negative) sentiment.
Critically, both reduce gender bias and political leaning bias
on Twitter/X (\textit{engaging}: z = $-$0.5, $-$0.6; \textit{popular}: z = $-$1.5,
$-$1.7), suggesting engagement-driven optimization partially mitigates demographic biases.


\subsection{Linguistic and Content Feature Bias by Dataset}
\label{sec:linguistic-bias}

\begin{figure}[h!]
  \centering
    \includegraphics[width=\linewidth]{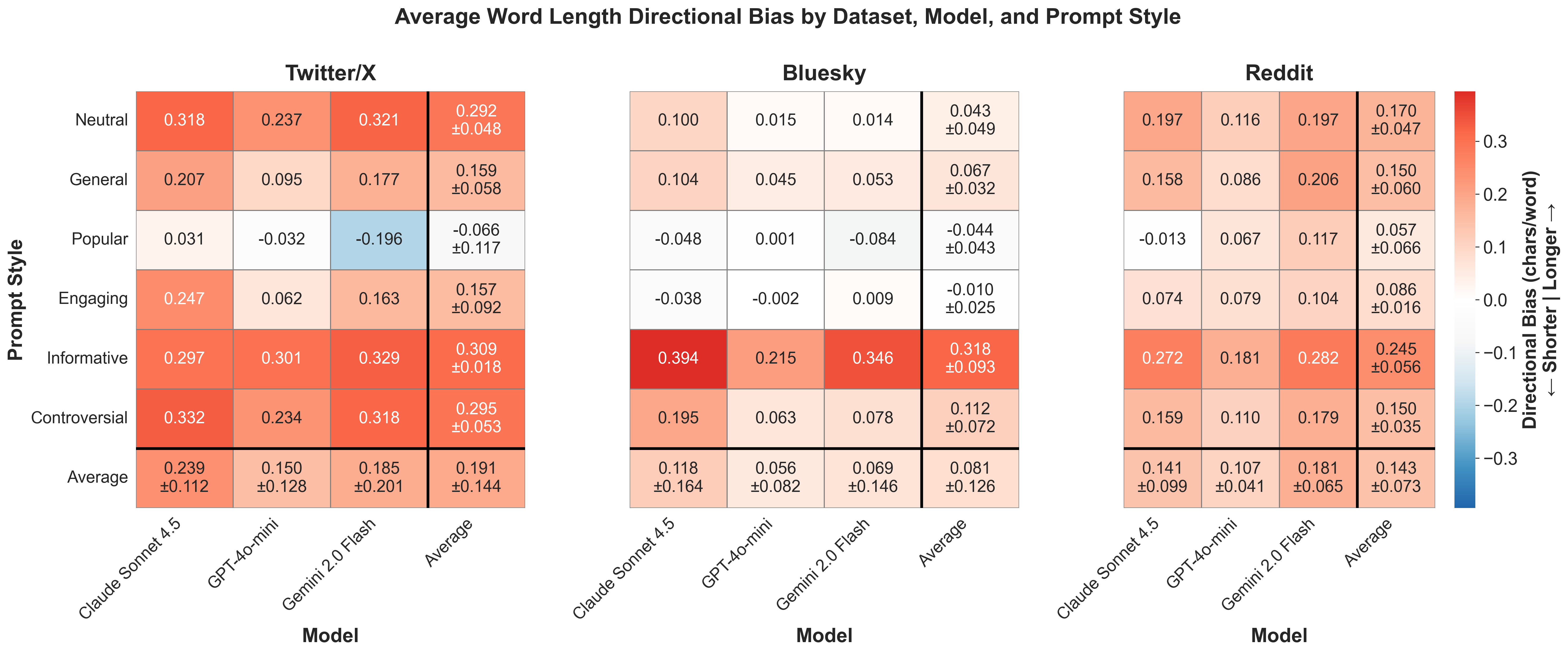}
  \caption{Average word length directional bias by dataset, model, and prompt style.
    A shared colour scale is used across the three dataset panels within each
    subfigure. Average rows/columns report mean\,$\pm$\,s.d.}
  \label{fig:09text}
\end{figure}

\begin{figure}[h]
  \centering
  \begin{subfigure}{\linewidth}
    \includegraphics[width=\linewidth]{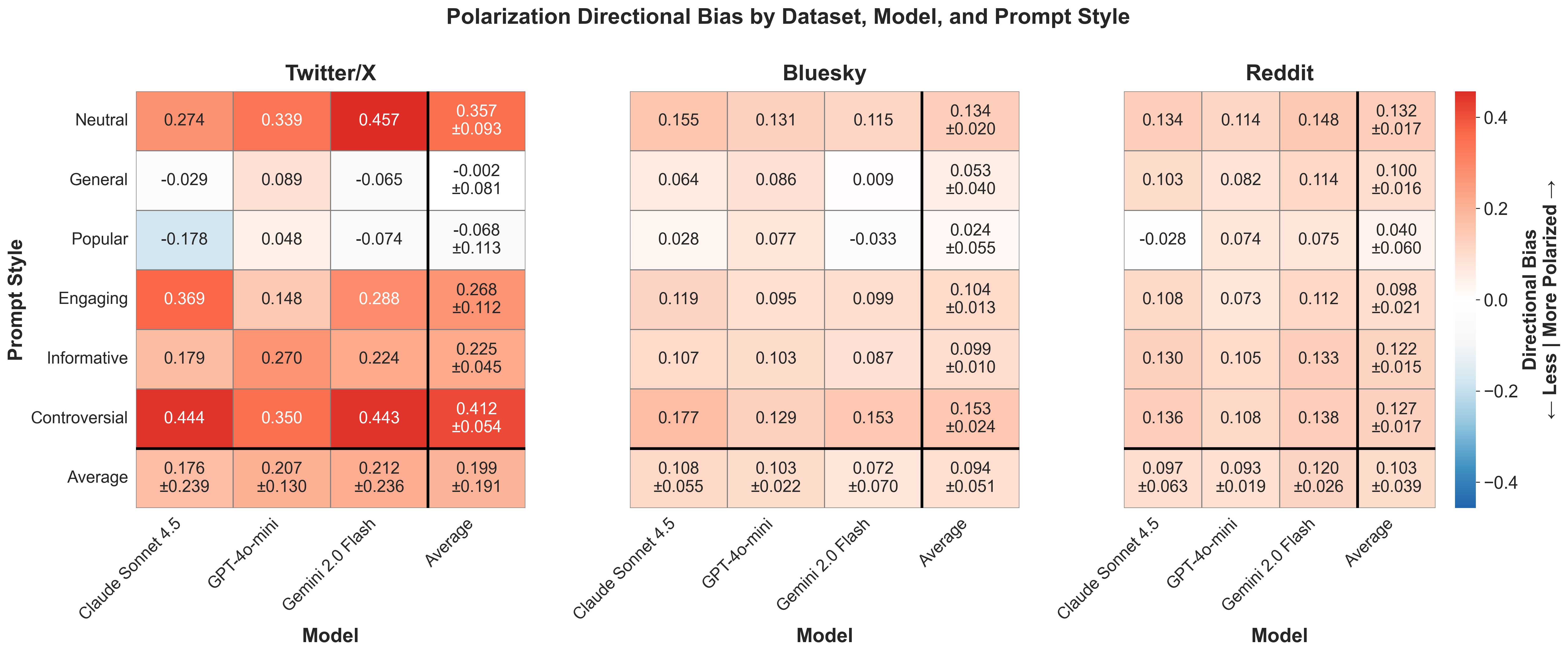}
    \caption{Polarization score. 
      }
    \label{fig:09c}
  \end{subfigure}
  \begin{subfigure}{\linewidth}
    \includegraphics[width=\linewidth]{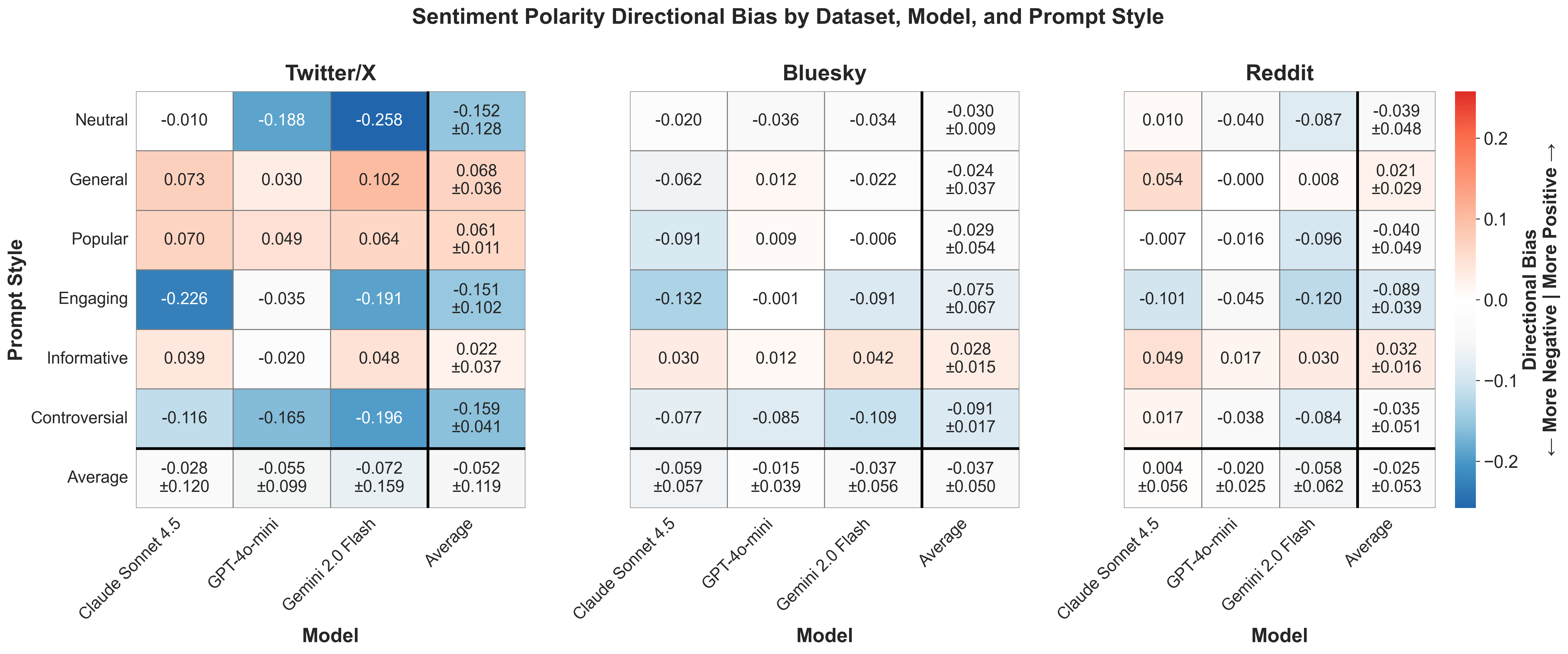}
    \caption{Sentiment polarity. 
     }
    \label{fig:09d}
  \end{subfigure}
  \begin{subfigure}{\linewidth}
    \includegraphics[width=\linewidth]{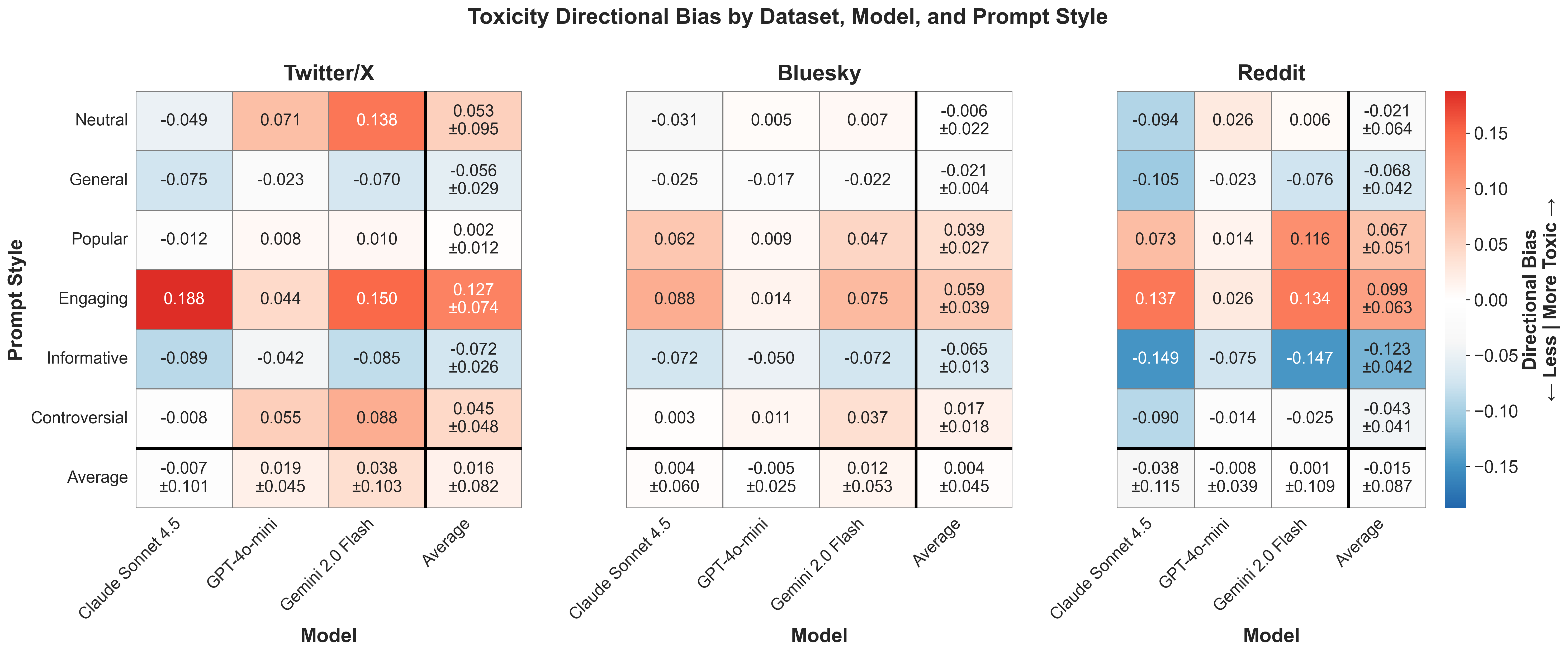}
    \caption{Toxicity. 
    }
    \label{fig:09e}
  \end{subfigure}
  \caption{Content-feature directional bias by dataset, model, and prompt style.
    All three features show strong prompt-style dependence.}
  \label{fig:09content}
\end{figure}

Figures~\ref{fig:09text}--\ref{fig:09content} show the raw directional bias for four
continuous features — average word length, polarization score,
sentiment polarity, and toxicity — broken down by dataset (subplots) and by
model $\times$ prompt-style combination (heatmap rows and columns).
All dataset panels within each figure share a common colour scale.
Average rows and columns show the mean $\pm$\,s.d.\ across prompt styles and
models respectively.

\paragraph{Average word length (Figure~\ref{fig:09text}).}
Recommended posts use slightly longer words on all three platforms
(Twitter/X: $\mu = +0.191$ chars/word; Reddit: $+0.143$; Bluesky: $+0.081$),
suggesting a consistent preference for more formal or technical vocabulary. 
The effect is the most pronounced under \emph{informative} prompt-styles. Conversely, \emph{popular} prompt-style leads almost consistently to slightly shorter words.

\paragraph{Polarization, sentiment, and toxicity (Figure~\ref{fig:09content}).}
Recommended content is more polarized than the pool on all platforms, with the
effect strongest on Twitter/X ($\mu = +0.199$) and weaker on Reddit ($+0.103$)
and Bluesky ($+0.094$).
The effect is stronger under \textit{controversial}, \textit{engaging}, but also \textit{neutral} prompt-style.
Recommended posts are on average slightly more negative in sentiment than the
pool, and the effect is stronger on Twitter/X and, again, with \textit{controversial}, \textit {engaging}, but also \textit{neutral} prompt-style. Some positive sentiment bias appears with \textit{general} and \textit{popular}, mostly on Twitter/X, though.
Toxicity bias is small on average ($\mu = +0.002$) but highly prompt- and platform-dependent:
the \emph{engaging} prompt raises toxicity ($\mu = +0.095$; max
Claude/engaging/Twitter/X $= +0.188$), while the \emph{informative} prompt suppresses it
($\mu = -0.087$, min Claude/informative/Reddit $= -0.149$).

\subsection{Topic-Level Directional Bias}
\label{sec:topic-bias}

\begin{figure}[h!]
  \centering
  \begin{subfigure}{\linewidth}
    \includegraphics[width=\linewidth]{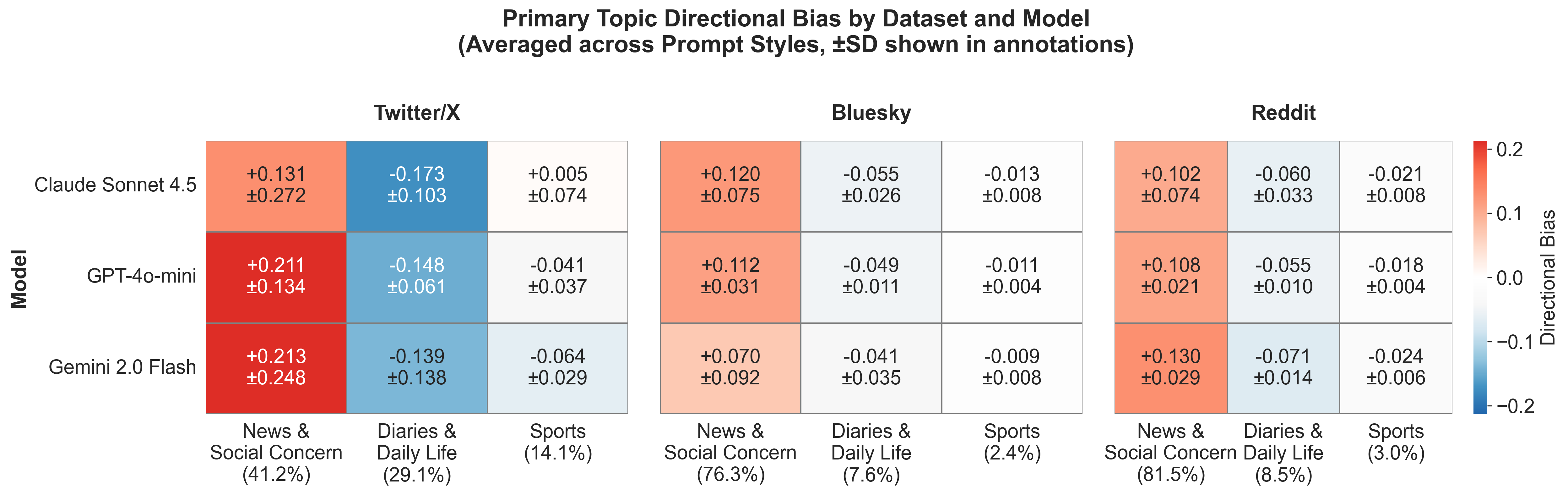}
    \caption{By model (averaged across prompt styles).}
    \label{fig:08a}
  \end{subfigure}
  \vspace{1em}
  \begin{subfigure}{\linewidth}
    \includegraphics[width=\linewidth]{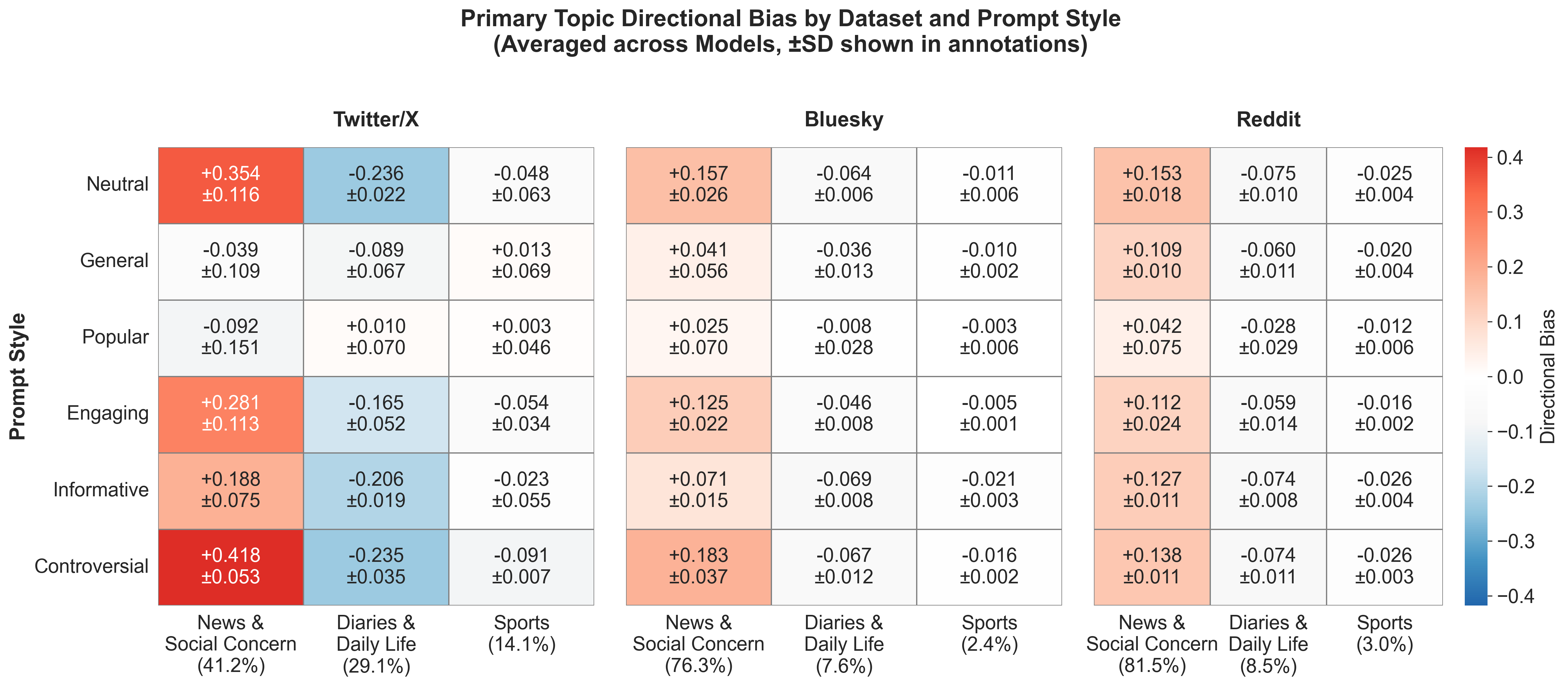}
    \caption{By prompt style (averaged across models).}
    \label{fig:08b}
  \end{subfigure}
  \caption{Primary-topic directional bias by dataset. Each cell shows the mean
    over-/under-representation of a topic relative to the candidate pool.
    Colors follow a diverging scale (blue = under-represented, red = over-represented).
    \emph{News \& Social Concern} is consistently the most over-represented topic
    (max $+0.418$; Twitter/X, controversial prompt), while \emph{Diaries \&
    Daily Life} is the most under-represented ($-0.236$; Twitter/X,
    neutral prompt).}
  \label{fig:08}
\end{figure}

Figures~\ref{fig:08a} and~\ref{fig:08b} disaggregate directional bias by the primary
topic of each post, using the three most frequent topics per dataset.
Across all datasets, \emph{News \& Social Concern} is the most consistently
over-represented topic, reaching a maximum directional bias of $+0.418$
(Twitter/X, controversial prompt).
\emph{Diaries \& Daily Life} shows the strongest negative bias ($-0.236$;
Twitter/X, neutral prompt), suggesting that personal and lifestyle
content is systematically under-recommended regardless of model.
Topic-level patterns are most pronounced on Twitter/X and weakest on Reddit and Bluesky.
The \emph{neutral} and \emph{controversial} prompt styles produce the largest
topic-level deviations, while \emph{popular} and \emph{general} prompts yield
distributions closer to the pool baseline.
The asymmetry between prompt styles is most visible for News \& Social Concern,
which is amplified under neutral and controversial prompts but suppressed under
popular prompts across all three datasets.

\subsection{Demographic Bias by Model and Prompt Style}
\label{sec:demo-bias-prompt}

\begin{figure}[h!]
  \centering
  \begin{subfigure}{\linewidth}
    \includegraphics[width=\linewidth]{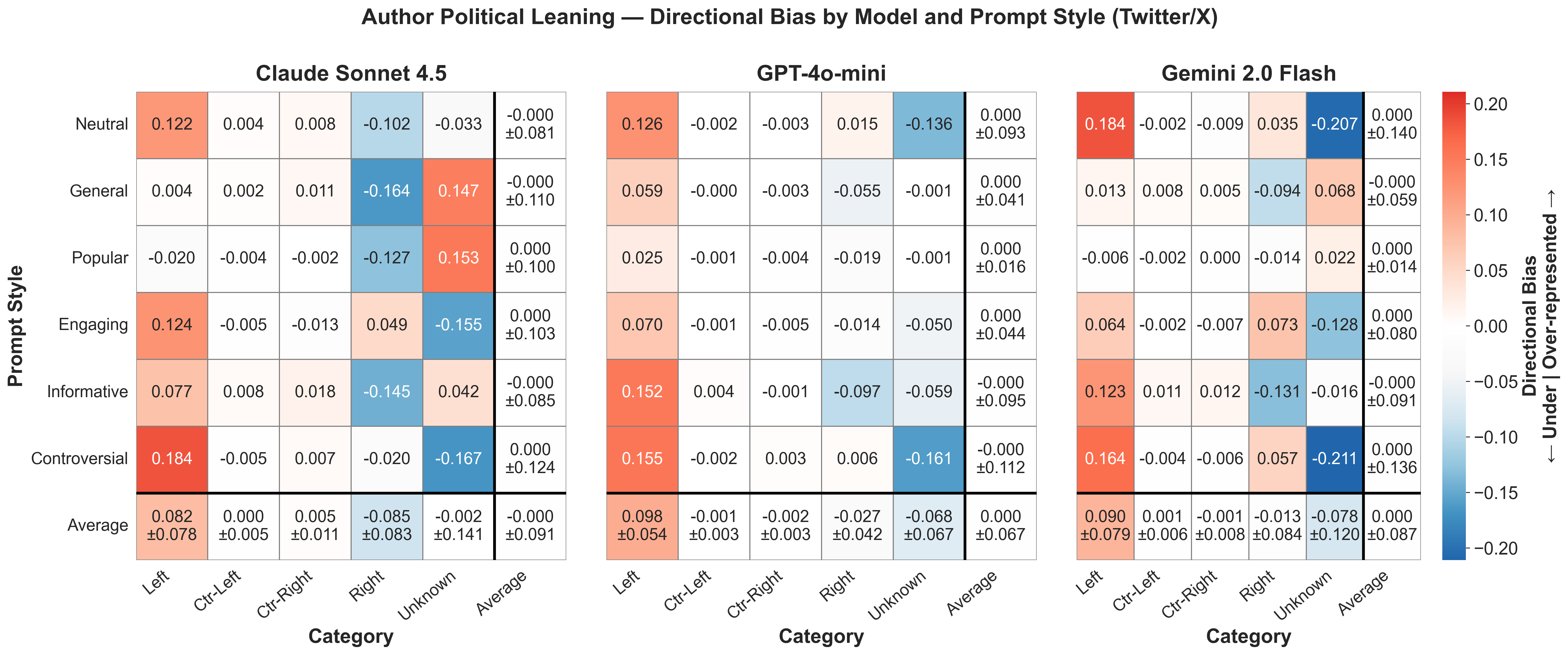}
    \caption{Political leaning. All models over-represent left-leaning authors
      ($17.9\%$ of pool) and under-represent right-leaning authors ($43.4\%$
      of pool).}
    \label{fig:10a}
  \end{subfigure}
  \vspace{0.8em}
  \begin{subfigure}{\linewidth}
    \includegraphics[width=\linewidth]{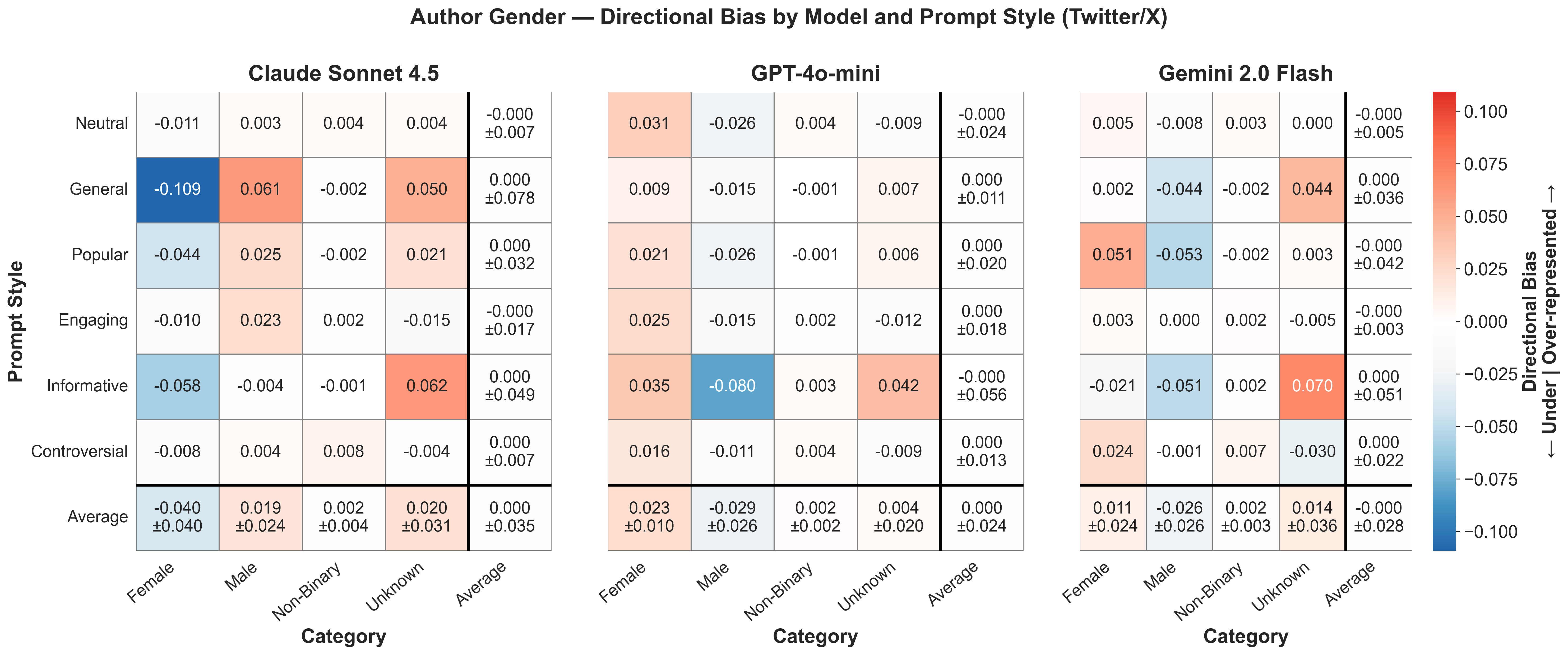}
    \caption{Gender. The direction of bias differs across models; Claude
      disfavours female authors while GPT-4o-mini and Gemini show small
      positive biases.}
    \label{fig:10b}
  \end{subfigure}
  \vspace{0.8em}
  \begin{subfigure}{\linewidth}
    \includegraphics[width=\linewidth]{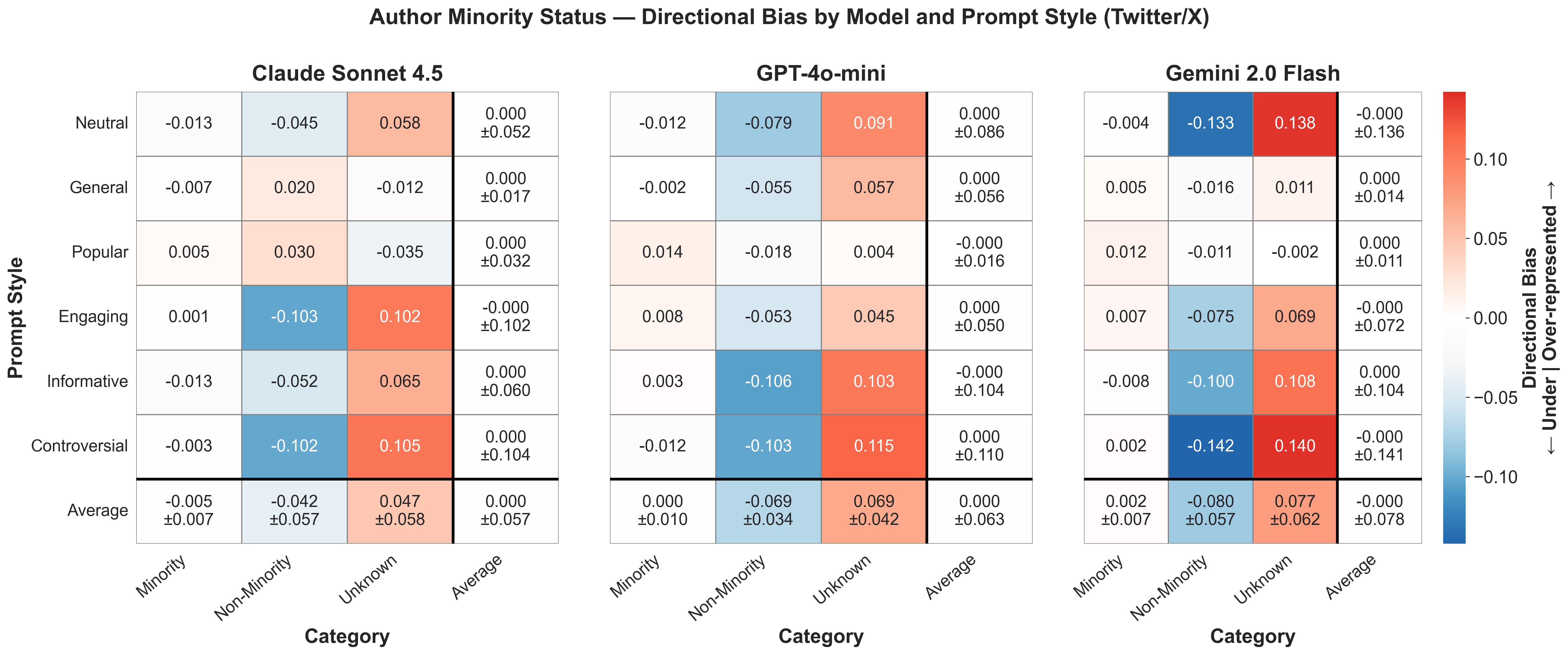}
    \caption{Minority status. All models under-represent the non-minority
      category and over-represent unknown-status authors; strongest for Gemini.}
    \label{fig:10c}
  \end{subfigure}
  \caption{Demographic directional bias by model and prompt style (Twitter/X only;
    LLM-inferred demographics). Each row is a prompt style; each column is a
    demographic category (pool proportions in parentheses). A shared colour
    scale is used across the three model panels within each subfigure.
    Average cell annotations report mean\,$\pm$\,s.d. that reflects variability
    across prompt styles.}
  \label{fig:10}
\end{figure}

Figure~\ref{fig:10} decomposes demographic directional bias (Twitter/X only;
LLM-inferred from user bios) along the prompt-style dimension, separately for
each model.
This complements Figure~\ref{fig:demographic_bias} in the main text, which averaged over prompt styles,
by revealing how strongly prompt wording modulates each form of demographic bias.
Bias values are zero-sum-normalised within each (model, prompt style) condition.

\paragraph{Political leaning (Figure~\ref{fig:10a}).}
Prompt style is a strong moderator of political bias.
The \emph{controversial} prompt produces the largest left-leaning over-representation
across all three models (mean $+0.167$; individual model values: Claude $+0.184$,
GPT-4o-mini $+0.155$, Gemini $+0.164$), consistent with models associating
controversy with politically charged. 
The \emph{neutral} prompt also yields a substantial left bias ($+0.144$).
By contrast, the \emph{popular} prompt nearly eliminates the left-leaning advantage
(mean $-0.001$) and produces the least skewed distribution across categories.
The right-leaning category is most suppressed under the \emph{informative} and
\emph{general} prompt styles (means of $-0.124$ and $-0.105$ respectively), whereas
under \emph{engaging} it is slightly over-represented ($+0.036$) — the only
condition in which right-leaning authors are not penalised on average.

\paragraph{Gender (Figure~\ref{fig:10b}).}
Gender bias is smaller in magnitude overall and shows a clearer prompt-style
dependence than the model-averaged results suggest.
The \emph{general} prompt is the most disadvantageous for female authors
(mean $-0.033$), while \emph{controversial} and \emph{popular} produce a
small positive bias ($+0.011$ and $+0.009$ respectively).
The \emph{informative} prompt strongly suppresses male authors ($-0.045$) while
boosting unknown-gender authors ($+0.058$).
The sign of female bias therefore flips depending on prompt: negative under
\emph{general} and \emph{informative}, near-zero or positive under
\emph{controversial}, \emph{engaging}, \emph{neutral}, and \emph{popular}.

\paragraph{Minority status (Figure~\ref{fig:10c}).}
The prompt-style effect on minority-status bias is concentrated in the
\emph{non-minority} and \emph{unknown} categories; the \emph{minority} (yes)
category remains near-zero across all conditions (range: $-0.010$ to $+0.010$).
The \emph{controversial} prompt produces the strongest suppression of non-minority
authors (mean $-0.116$) paired with the largest boost to unknown-status authors
($+0.120$).
The \emph{popular} prompt is again the most neutral, yielding a near-zero bias
for non-minority authors ($0.000$) and a slight negative shift for unknown
($-0.011$) — the only condition where this asymmetry reverses.

\subsection{Feature Importance and Bias Mechanisms}
\label{sec:feature_analysis}
To understand which features drive decisions,
we train Random Forest classifiers (100 trees, max depth 10, balanced class weights, min 10
samples/leaf) predicting recommendation status for each of the 54 conditions, assessed via
5-fold cross-validation mean AUROC is 0.746 (range: 0.571–0.918). SHAP values~\cite{lundberg2017unified} quantify per-feature contributions; aggregate importance
for feature $j$ is:
\begin{equation*}
\text{SHAP}_j = \frac{1}{n}\sum_{i=1}^{n} |\text{SHAP}_j(x_i)|
\end{equation*}

\begin{figure*}[h]
\centering
\includegraphics[width=\textwidth]{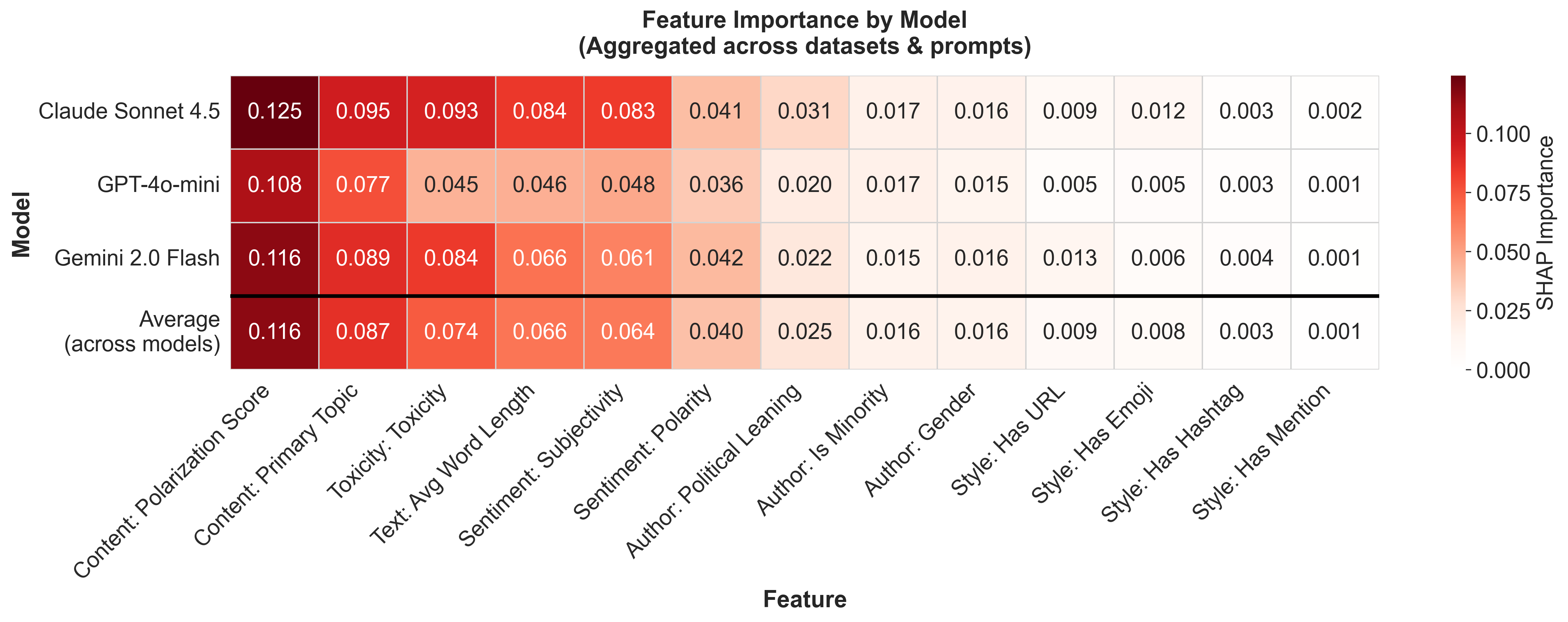}
\caption{\textbf{Feature importance by model.} SHAP importance values aggregated across
datasets and prompts. Features ordered by decreasing average importance (left to right);
bottom row shows cross-model averages. Claude shows the highest polarization and toxicity
reliance; GPT-4o-mini exhibits the most
balanced feature usage.}
\label{fig:importance-by-model}
\end{figure*}

Figure~\ref{fig:importance-by-model} reveals the mechanisms behind these biases through
SHAP feature importance. Polarization dominates all models (Claude: 0.125, Gemini: 0.116,
GPT-4o-mini: 0.108). 
Primary topic and Toxicity rank second and third overall (0.087 and 0.074 on average). 
Crucially, demographic features show near-zero direct importance (SHAP $<$ 0.03) despite producing measurable bias, confirming \textit{indirect discrimination}: demographic bias emerges through learned correlations with
content features rather than explicit reliance on author attributes. This mechanism resists
simple mitigation --- masking demographic features would not eliminate the bias.

\end{document}